\documentclass[twocolumn,aps,prl]{revtex4}
\usepackage[english]{babel}
\usepackage[dvipdf]{graphicx}
\usepackage{amssymb}
\usepackage{amsmath}
\usepackage{amsbsy}

\usepackage{dcolumn}% Align table columns on decimal point

\usepackage{color}
\begin{document}

\title{
  Structural signature of a brittle-to-ductile transition in self-assembled networks}

\author{Laurence Ramos$^{1,2,\star}$, Arnaud Laperrousaz$^{1,2}$, Philippe Dieudonn\'{e}$^{1,2}$,
and Christian Ligoure$^{1,2}$}

\affiliation{
$^1$Universit\'{e} Montpellier 2, Laboratoire Charles Coulomb UMR 5221, F-34095, Montpellier,
France\\
$^2$CNRS, Laboratoire Charles Coulomb UMR 5221, F-34095, Montpellier, France}

%{\small* Email: neda.ghofraniha@univ-montp2.fr}
\begin{abstract}
We study the nonlinear rheology of a novel class of transient networks, made of surfactant micelles of tunable morphology reversibly linked by block copolymers. We couple rheology and time-resolved structural measurements, using synchrotron radiation, to characterize the highly nonlinear viscoelastic regime. We propose the fluctuations of the degree of alignment of the micelles under shear as a probe to identify a fracture process. We show a clear signature of a brittle-to-ductile transition in transient gels, as the morphology of the micelles varies, and provide a parallel between the fracture of solids and the fracture under shear of viscoelastic fluids.
\end{abstract}

\maketitle

Brittle and ductile fractures of solids have been extensively studied \cite{Richie1999,Ravi2000,Lewandowski2005,Picallo2010}. In this field, recent experiments on various materials ranging from bones to carbon nanotubes and gels \cite{Nardelli1998,Peterlik2006,Baumberger2006} aim at a characterization of the microscopic mechanisms at play in the brittle-to-ductile transition. Soft solids such as hydrogels \cite{Itagaki2010} or foams \cite{Arif2010} can also exhibit brittle-to-ductile transitions. Referring to fracture in transient networks is more delicate as this class of materials is viscoelastic and intrinsically self-healing. In this case, whether fractures are essentially observed when the transient network is solicited at a shear rate larger than the inverse of its intrinsic relaxation time \cite{Berret2001,Skrzeszewska2010,Olsson2010} remains an open question. Experiments have nevertheless demonstrated the brittle character of a fracture for viscoelastic fluids made of microemulsion droplets reversibly linked by copolymers \cite{Tabuteau2009}. On the other hand, simulations have shown a transition in transient viscoelastic gels from brittlelike fracture to ductilelike fracture \cite{Sprakel2009}. However, experimental evidence of a brittle-to-ductile transition in transient networks is still lacking.

We have recently designed a novel class of transient self-assembled networks by reversibly bridging surfactant micelles with telechelic block copolymers. By changing in a controlled fashion the morphology of the micelles (from sphere to rod to flexible worm), both the linear viscoelasticity and the nonlinear rheology of the networks can be tuned \cite{Tabuteau2008,Tixier2010}. The shear stress vs shear-rate flow curves for samples comprising short micelles display features characteristic of a viscoelastic material undergoing a brittle fracture \cite{Tabuteau2009,Tixier2010}, whereas those of samples comprising long and entangled micelles show a typical shear-banding process. In order to characterize the transition between the two regimes, we couple rheology and time-resolved structural measurements, using synchrotron radiation. We show how a combination of these two techniques provides a clear signature of a brittle-to-ductile transition in viscoelastic fluids, as the morphology of the micelles varies, and provide arguments for drawing a parallel between the fracture of solids and the fracture of viscoelastic fluids under shear.

%SAMPLE DESCRIPTION+METHODS

\begin{figure}
\includegraphics[width=8cm]{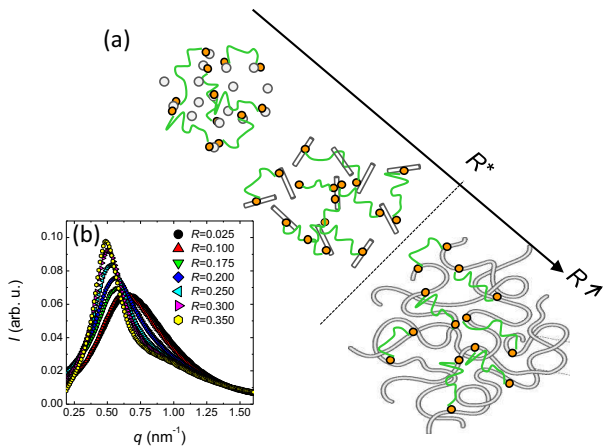}
\caption{ (a) Cartoon of the transient networks with tunable morphologies; $R^*=0.225$ is defined as the critical value of the cosurfactant over surfactant molar ratio, $R$, above which micelles are entangled; (b) Scattered intensity versus $q$ vector for samples with various $R$, as indicated in the legend. }
\label{fig:1}
\end{figure}

Samples description is given elsewhere \cite{Tixier2010}. In brief, we dilute a surfactant (cetylpyridinium chloride) in an aqueous solvent, made of $12$ wt$\%$ of glucose and  $88$ wt$\%$ wt water, with a NaCl concentration of $0.5$ M. We add a controlled amount of sodium salicylate as cosurfactant, in order to tune the morphology of the micelles from sphere to rod to worm, as the co-surfactant over surfactant molar ratio, $R$, increases  \cite{Rehage1988}.
Micelles are reversibly linked by triblock copolymers (see fig.~\ref{fig:1}a for a cartoon of the structures). The polymer used is a home-synthesized hydrophobically modified water-soluble poly(ethylene oxide) (PEO) ($M_{\rm{w}}=10 000$ g/mol), with C$_{23}$H$_{47}$ groups grafted to each extremity of the PEO chain.
We fix the mass fraction of surfactant at  $9\%$, and the mass ratio of polymer over surfactant plus cosurfactant at $55\%$, and vary $R$ between $0$ and $0.35$.
Samples are prepared by dissolving the surfactant, co-surfactant, and copolymer, in the aqueous solvent. After a few days at rest, all samples are one-phase transparent and homogeneous mixtures.

Scattering experiments have been performed on the ID-2 beam line at ESRF,
Grenoble, France (preliminary experiments were performed on the SWING beam line at Soleil). A stress-controlled Haake RS300 rheometer
equipped with a Couette cell is used online. The incident beam is radial with respect to the cell and the two-dimensional (2D) scattering patterns are collected in the flow-vorticity plane. In a typical experiments, the sample is submitted to a constant shear-rate, $\dot{\gamma}$, and images (typically $100$) of the scattered intensity are taken at regular intervals (typically every $2$ or $3$ s). Exposure time is fixed at $0.1$ s. The height at which the incident beam shines the sample in the Couette cell is systematically changed between two measurements in order to avoid radiation damage. Temperature is fixed at $23^\circ$C.

%RESULTS STRUCTURE

Figure~\ref{fig:1}b shows the azimuthally averaged scattered intensity, $I$, versus wave vector, $q$, for several samples with different $R$. The local structure is not affected by the shear. All spectra exhibit a broad peak, which becomes sharper when $R$ increases (Fig.~\ref{fig:1}b) and which originates from the steric repulsion between the micelles induced by the polymers \cite{Massiera2002,Massiera2003,Ramos2007}.
The peak position, $q^*$, is related to the average distance between micelles and can be linked to the average length of the micelles. The decrease of $q^*$ when $R$ increases is a consequence of the elongation of the micelles with $R$, but a quantitative determination of the average micelle length is difficult due to the very high (exponential) polydispersity  of cylindrical micelles \cite{Cates1990}.

%-----------------------------------  FIG1 -------------------------------------
%------------------------------------ FIG1 --------------------------------------

%RESULTS VISCO-ELASTICITY AND FLOW CURVES

\begin{figure}
\includegraphics[width=8cm]{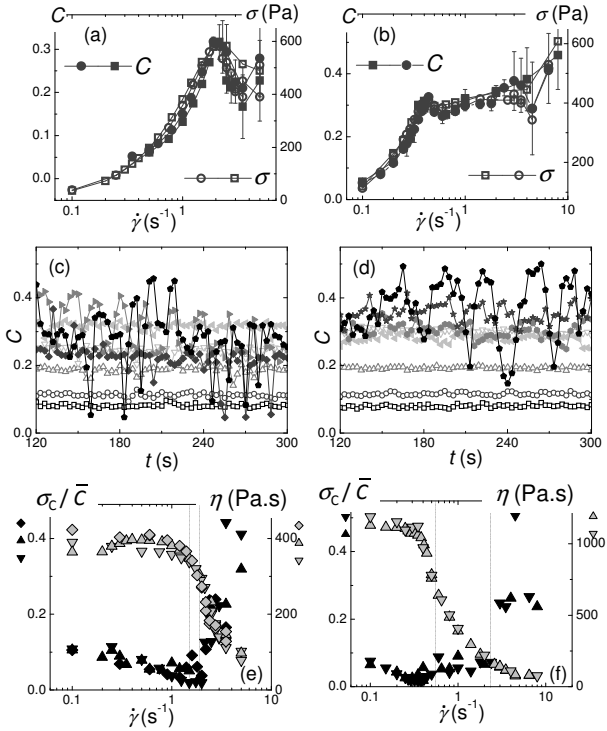}
\caption{(a,b) Shear stress (full symbols) and contrast (open symbols) as a function of the shear
rate applied, $\dot{\gamma}$, in the steady state. The two sets of data in (a) and (b) correspond to two
independent measurements.
The bars for the contrast data correspond to the standard deviation.
(c,d) Time evolution of the contrast in the steady state as a function of $\dot{\gamma}$.
(e,f) Normalized fluctuations of the contrast (black symbols) and viscosity (gray symbols)
 as a function of $\dot{\gamma}$. In (a,c,e) $R=0.30$
and in (b,d,f) $R=0.35$.
}
\label{fig:3}
\end{figure}

Consistently with our previous results \cite{Tixier2010}, the linear viscoelasticity  of the self-assembled networks is well described by a one mode Maxwell fluid model for $0 \leq R  \leq 0.2$ (with an elastic plateau and a characteristic time in the range $5600-7600$ Pa, and  $12-46$ ms, respectively), when the micelles are too short to be entangled,
and by a two-mode Maxwell fluid model \cite{Nakaya2008} for $R \geq 0.25$, when the micelles are sufficiently long and entangled, with elastic plateau, $G_{\rm{fast}}$ (respectively $G_{\rm{slow}}$) in the range $2200-3000$ Pa (respectively $1200-2000$ Pa) and characteristic time, $\tau_{\rm{fast}}$ (resp. $\tau_{\rm{slow}}$)  in the range $22-90$ msec (resp. $80-1100$ msec) for the fast (resp. slow) mode.  We define $R^*=0.225$ as the crossover between the two regimes (fig.~\ref{fig:1}a). Flow curves, stress  $\sigma$ \textit{vs} shear-rate $\dot{\gamma}$,  are measured online, while probing the structure. Two markedly different types of flow curves are observed at low and high $R$.  For $R\leq 0.3$, an abrupt drop of $\sigma$ is  measured above a critical $\dot{\gamma}$ (fig.~\ref{fig:3}a). Previous results showed that the drop of the stress is accompanied by a drop of the first normal stress difference, $N$, and signs a fracture mechanism \cite{Tixier2010}. On the other hand,
for a sample with a slightly larger $R$ ($R=0.35$) a plateaulike variation
 of  $\sigma$ is measured above a critical $\dot{\gamma}$ (Fig.~\ref{fig:3}b), while $N$ steadily increases, signing shear-banding, a well documented  mechanism observed in solutions of entangled wormlike micelles \cite{Lerouge2010}. Note that the transition between the two types of behavior at high shear rates seems sharper in this work than in our previous work \cite{Tixier2010}. This is presumably due to differences in the geometries used (here a Couette cell with smooth surface,  whereas
 a plate-plate geometry with rough surfaces was used previously).

%COMPARISON SAXS / RHEOLOGY

\begin{figure}
\includegraphics[width=8cm]{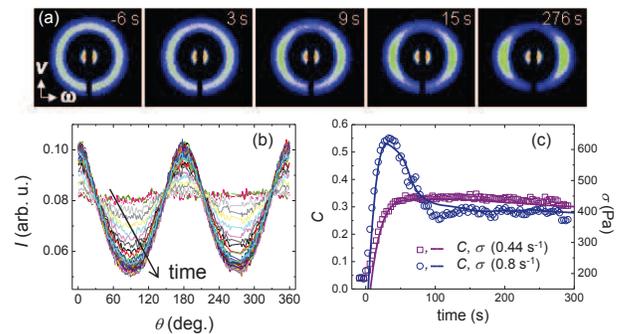}
\caption{(a) Time series of 2D scattering patterns for a sample with $R=0.35$
and a shear rate of $0.44 \rm{s}^{-1}$. The indicated time corresponds to
 the time elapsed since the application of the shear ; time evolution of (b) the
 azimuthal profiles measured at the peak positions and (c) the shear stress (line)
 and the contrast (squares), as measured from profiles as those shown in (b).}
\label{fig:2}
\end{figure}

We always find 2D isotropic scattering patterns for samples with $R<0.175$, independently of the shear rate
applied. This indicates that the micelles are randomly oriented in solution, presumably because they are too short to align sufficiently under flow. These samples were not analyzed further.
For samples with longer micelles, we find by contrast that the anisotropy of the pattern
develops with time when submitted to a fixed $\dot{\gamma}$. A time series of 2D patterns is shown in fig.~\ref{fig:2}a.  The higher intensity in the direction perpendicular
to the velocity direction (at an azimuthal angle $\theta_{max} \simeq 180$ deg) indicates that the micelles are preferentially aligned along the direction of the imposed flow.
In order to quantify
the degree of alignment
we compute the azimuthal profile, taken at the peak position. A series of profiles measured
at regular intervals after application of a shear rate are shown in fig.~\ref{fig:2}b.
We calculate the contrast of the profiles, defined as
 $C=(I_{\rm{max}} - I_{\rm{min}}) / (I_{\rm{max}} + I_{\rm{min}})$,
where $I_{\rm{max}}$ and $I_{\rm{min}}$ are respectively the maximal and minimal intensities
measured along the profile. By construction $C$ varies between $0$, for an isotropic profile, and $1$,
for highly anisotropic profiles. In practice, we found that in our experimental conditions, $C \cong 0.05$ for
an isotropic signal, due to the experimental noise.

%----------------------------------- FIG2 -------------------------------------
%------------------------------------FIG2--------------------------------------

%-----------------------------------  FIG3-------------------------------------
%------------------------------------FIG3--------------------------------------

\begin{figure}
\includegraphics[width=8cm]{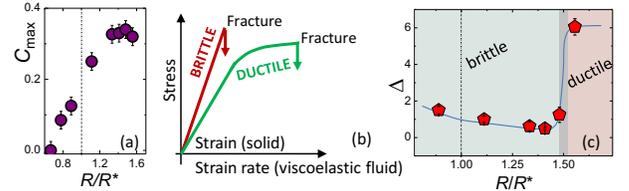}
\caption{(a) Contrast at the onset of stress plateau or stress drop as a function of the cosurfactant over surfactant molar ratio, $R$, normalized by $R^*$, where $R^*=0.225$ is defined as the critical value of $R$ above which micelles are entangled. (b) Connection between solids and viscoelastic fluids. (c) Relative difference between shear rates for fracture and nonlinearity, $\Delta$, as defined in the text, as a function of $R/R^*$.
}
\label{fig:4}
\end{figure}

Remarkably, we found that upon application of a fixed $\dot{\gamma}$, the time evolution of $C$
follows that of the stress, as illustrated in fig.~\ref{fig:2}c for two different $\dot{\gamma}$. When $\sigma$ smoothly increases to its steady state value, so does the contrast. On the other hand, when a stress overshoot is measured, an overshoot of the contrast is measured concomitantly. The mapping between $C$ and $\sigma$ recalls the stress optical law, which relates the stress to the refractive index tensors of a material, even if it does not apply here as scattering data are not collected in the appropriate (flow-gradient) plane \cite{Shikata1994,Helgeson2009}.
The mapping between $\sigma$ and $C$ is also observed when considering the
steady state value of the two parameters for a given $\dot{\gamma}$. This is
illustrated in fig.~\ref{fig:3}a,b where the stationary values for $C$ and $\sigma$ are plotted as a function of the shear rate applied, for two markedly different flow curves, obtained for a sample with $R=0.30$ ($G_{\rm{fast}}=2430$ Pa, $\tau_{\rm{fast}}=25$ ms, $G_{\rm{slow}}=1600$ Pa, $\tau_{\rm{slow}}=172$ ms), and for a sample with $R=0.35$ ($G_{\rm{fast}}=2200$ Pa, $\tau_{\rm{fast}}=105$ ms, $G_{\rm{slow}}=1270$ Pa, $\tau_{\rm{slow}}=735$ ms).
We find that the maximum contrast reached at the onset of stress drop or stress plateau increases continuously when $R$ increases and the micelles becomes longer, as expected intuitively, and reaches a plateau of the order of $C_{\rm{max}} \simeq 0.3$ for $R \geq 0.3$ ($R/R^* \geq 1.3$) (fig.~\ref{fig:4}a). Clearly this quantity cannot be used to characterize the transition evidenced in the nonlinear behavior of the networks as the two samples studied in fig.~\ref{fig:3} display markedly qualitatively different flow curves but have equal $C_{\rm{max}}$. Instead, we show below that the fluctuations of the contrast \cite{Angelico2010} appear as a good parameter to apprehend the transition.

%ANALYSIS OF THE CONTRAST FLUCTUATIONS

Time evolutions of the contrast for several $\dot{\gamma}$
are plotted in fig.~\ref{fig:3}c,d, for the same samples as in fig.~\ref{fig:3}a,b. Data are displayed in a time-window
corresponding to a steady state, for which $C$ is on average constant, for all $\dot{\gamma}$.
As described above, the average value is strongly dependent on $\dot{\gamma}$. More surprisingly,
we also find that the fluctuations of the contrast strongly depend on $\dot{\gamma}$. At low $\dot{\gamma}$,
fluctuations are weak, while they are much more important at higher $\dot{\gamma}$.
In order to quantify the contrast fluctuations, we compute $\sigma_ C / \overline{C}$, where $\sigma_ C$ and $\overline{C}$ are the standard deviation and the time-averaged value of $C$, respectively. At low $\dot{\gamma}$, $\sigma_ C / \overline{C}$ is small (typically between $2$ and $10$ \%), whereas it can reach $50$ \% at high $\dot{\gamma}$.
Figure~\ref{fig:3}e,f display $\sigma_ C / \overline{C}$ as
a function of $\dot{\gamma}$, for the samples with $R=0.3$ and $R=0.35$.  Note that data acquired and analyzed for several samples with a given $R$ nicely overlap, demonstrating the reproducibility of the data and the robustness of the analysis. Interestingly, the transition from a regime when fluctuations are weak to a regime where the fluctuations are strong is rather sharp. This enables us to define a critical shear rate for the onset of strong fluctuations, $\dot{\gamma}_{\rm{fracture}}$.
We consider the onset of strong fluctuations as a signature of the fracture of the networks. Indeed, as expected for a fracture mechanism, the transition is abrupt. Moreover,
strong fluctuations of the contrast are correlated with strong fluctuations of the position of the maximum intensity in the azimuthal profile, $\theta_{max}$: the standard deviation of $\theta_{max}$ increases abruptly from  $4-5$ deg. in the Newtonian regime to values between $10$ and $20$ deg. when the fluctuations of the contrast increase, as expected if blocks of materials comprising aligned micelles rotate independently in the gap of the Couette cell. In addition, for a sample with a structure very close to that of the samples investigated here before micellar entanglements ($R<R^*$), a fracture mechanism has been visualized at the stress drop, which occurs concomitantly with the onset of strong fluctuations \cite{Tabuteau2009}, hence reenforcing the connection between fracture and contrast fluctuations.

%-----------------------------------  FIG4-------------------------------------
%------------------------------------FIG4--------------------------------------

In solid materials, brittle fracture occurs in the elastic regime, characterized by a straight line in a stress \textit{vs} strain plot (Fig.~\ref{fig:4}b). By contrast, plastic deformation accumulates before fracture in ductile materials, which translates into a decrease of the local slope of the stress vs strain  \textit{vs} curves (fig.~\ref{fig:4}b). By analogy with solid materials, and mapping the strain in solids to the shear rate of viscoelastic fluids under shear, we therefore propose to compare the shear rate for the onset of fracture, $\dot{\gamma}_{fracture}$ and the shear rate for the onset of nonlinearity, $\dot{\gamma}_{thin}$. We define $\Delta$ as the normalized difference between the two critical shear rates
$ \Delta = (\dot{\gamma}_{fracture} - \dot{\gamma}_{thin}) / \dot{\gamma}_{thin}$. The parameter $\Delta$ introduced here is the equivalent to the amount of plastic deformation accumulated before fracture in ductile materials (fig.~\ref{fig:4}b). Experimentally, $\dot{\gamma}_{thin}$ is directly evaluated as the shear rates at which the samples is not Newtonian anymore (i.e. the onset of shear thinning).
Together with the data for the contrast fluctuations, we therefore plot the viscosity, $\eta$ as measured concomitantly (fig.~\ref{fig:3}e,f).
We observe that both strong fluctuations and shear-thinning (defined when $\eta$ has decreased by more than $10 \%$ of its Newtonian low-$\dot{\gamma}$ value) occurs at more or less the same shear rate for the sample with $R=0.30$ while onset of strong fluctuations and fracture occurs much later than shear thinning  for the sample with $R=0.35$.
We find that $\Delta$ is low, $\Delta = 1.0 \pm 0.4$, when the micelles are relatively short and shoots up very abruptly at $ \Delta =6.1 \pm 0.5$ when the micelles are slightly longer. This allows one to define a transition between "brittlelike" behavior and "ductilelike" behavior when the micelles elongate (fig.~\ref{fig:4}c).

What drives the transition from brittle to ductile in these materials is not clear yet. It is neither the ability of the micelles to align under shear nor the entanglements of the micelles as both samples with $R=0.30$ and with $R=0.35$ align in a similar fashion and are modeled by a double-Maxwell fluid, whereas their nonlinear rheologies are strikingly different. Instead our data show that micelles should be sufficiently long to exhibit a "ductile" behavior, suggesting connection with experimental observations on pure surfactant systems where a transition between shear thinning and shear banding is measured for sufficiently entangled wormlike micelles \cite{Berret1997,Hu2008}.

To conclude, we have investigated transient networks made of micelles whose morphology can be continuously tuned, and which are reversibly linked by polymers and have shown that the micelles can align under shear. We have proposed the fluctuations of the degree of alignment of the micelles as a structural probe of a fracture process of the materials under shear and have drawn an analogy between the fracture of solids and that of viscoelastic fluids. Depending on the morphology of the micelles, we found that the critical shear rate for fracture is either very close to the shear-rate at which the sample departs from its linear behavior (for "brittlelike" samples) or significantly larger than the shear rate at which the sample departs from its linear behavior (for "ductilelike" samples). To go further in the analogy between solids and viscoelastic fluids, we are currently studying the fracture propagation in the different types of networks.

We thank T. Phou for polymer synthesis, and J. Gummel at ESRF and F. Meneau at Soleil for technical assistance during the synchrotron experiments.
This work was supported in part by the ANR under Contract
No. ANR-06-BLAN-0097 (TSANET) and by the ESRF and Soleil synchrotron.

* laurence.ramos@univ-montp2.fr

%--------------------------------------------- REFERENCES -----------------------------------------
%\begin{thebibliography}{28}
%\end{thebibliography}

%\bibliographystyle{unsrt}
%\bibliographystyle{apsrev4-1.bst}

\end{document}